\documentclass[12pt]{article}
\input{epsf}
\usepackage{epsf,cite,epsfig,psfig,float,amssymb,stmaryrd,latexsym}
\usepackage{graphics,psfrag}
\usepackage{a4wide,graphicx}
\usepackage{cite}

\newcommand{\be}{\begin{equation}}
\newcommand{\ee}{\end{equation}}
\newcommand{\ba}{\begin{eqnarray}}
\newcommand{\ea}{\end{eqnarray}}
\newcommand{\vs}{\vspace{-0.275cm}}
\newcommand{\del}{\partial}
\newcommand{\rth}{\frac{1}{\sqrt{3}}}
\newcommand{\rsix}{\frac{1}{\sqrt{6}}}
\newcommand{\pr}{^\prime}
\newcommand{\Gm}{\Gamma}

\begin{document}

\title{Study of a possible $S=+1$ dynamically generated baryonic resonance}

\author{Sourav Sarkar, E. Oset and M.J. Vicente Vacas\\
{\small Departamento de F\'{\i}sica Te\'orica and IFIC,
Centro Mixto Universidad de Valencia-CSIC,} \\
{\small Institutos de
Investigaci\'on de Paterna, Aptd. 22085, 46071 Valencia, Spain}\\
}

\date{\today}

\maketitle
\begin{abstract}
Starting from the lowest order chiral Lagrangian for the interaction of the
baryon decuplet with the octet of pseudoscalar mesons we find an attractive
interaction in the $\Delta K$ channel with $L=0$ and $I=1$, while the interaction
is repulsive for $I=2$. The attractive interaction leads to a pole in the second
Riemann sheet of the complex plane and manifests itself in a large strength of the  $\Delta K$ scattering
amplitude close to the $\Delta K$ threshold, which is not the case for $I=2$.
However, we also make a study of uncertainties in the model and conclude that the
existence of this pole depends sensitively upon the input used and can disappear
within reasonable variations of the input parameters. We take advantage to study
the stability of the other poles obtained for the $\frac{3}{2}^-$  dynamically
generated resonances of the model and conclude that they are stable and not
contingent to reasonable changes in the input of the theory. 
\end{abstract}

\newpage

\section{Introduction}

   The dynamical generation of baryonic resonances within a
chiral unitary approach has experienced much progress from early works
which generated the $\Lambda(1405)$ \cite{Kaiser:1995eg,Oset:1997it}. 
Further studies unveiled other dynamically generated resonances which can be
associated to known resonances and others  found
new states \cite{oller,Oset:2001cn,Garcia-Recio:2002td}. Recently,
it has been shown that there are
actually two octets and a singlet of dynamically generated $J^ P=1/2^-$
resonances, which include among others two  $\Lambda(1405)$ states, the
$\Lambda(1670)$, the $\Sigma(1650)$ and a possible $I=1$ state close to the 
$K^- p$ threshold \cite{oller,Jido:2003cb,Garcia-Recio:2003ks}.

   What we call dynamically generated resonances are states which appear in a
natural way when studying the meson baryon interaction using coupled channel
Bethe-Salpeter equations (or equivalent unitary schemes) with a kernel
(potential) obtained from the lowest order chiral Lagrangian. This subtlety
is important since higher order Lagrangians sometimes contain information on
genuine resonances, and unitary schemes like the Inverse Amplitude Method
(IAM) \cite{dobado,iam} make them show up clearly, giving the appearance that they have been
generated dynamically, when in fact they were already hidden in the higher
order Lagrangian. This is the case of the vector mesons in the pseudoscalar
meson-meson interaction, which are accounted for in the ${\cal L}^{(4)}$ 
Lagrangian of Gasser and Leutwyler \cite{Gasser:1984gg} among other
interactions. This is shown in~\cite{derafael} where assuming explicit vector
meson exchange and also scalar meson exchange the values of the $L_i$
coefficients of \cite{Gasser:1984gg} can be reproduced. The introduction of the term genuine resonance,
as opposed to dynamically generated, finds its best definition within the
context of the large $N_c$ counting. In the limit of large $N_c$ there are
series of
resonances which appear \cite{derafael,schat} which we call genuine resonances. In this limit the
loops that characterize the series of the Bethe-Salpeter equation vanish and the
dynamically generated resonances fade away \cite{nsd}. The genuine
resonances cannot be generated dynamically and then this establishes a
distinction between the different resonances, the nature of which can be
distinguished when looking at the evolution of the poles as we gradually make
$N_c$ large. This exercise in the meson-meson interaction
\cite{nsd,pelaez} shows that
the $\sigma(500)$, $f_0(980)$, $a_0(980)$ scalar resonances are dynamically
generated and disappear in the large $N_c$ limit, while the $\rho$, $K^*$ remain
in this limit. These findings would not alter the philosophy of ref. 
\cite{derafael}, making the exchange of vector mesons and scalar mesons responsible for the
$L_i$ coefficients, but the choice of the particles exchanged, in the sense that the
scalar mesons to be used there should not be the lowest lying ones mentioned
above, but the nearest ones in energy in the Particle Data Book (PDB)\cite{PDB}. 

   Coming back to the meson baryon case, this distinction holds equally and there
are some resonances which are dynamically generated from the meson-baryon
interaction, solving the Bethe-Salpeter equations in coupled channels, while
there are others (the large majority) which do not qualify as such and stand,
hence, as genuine resonances. From the point of view of constituent quarks, the
latter ones would basically correspond to $3q$ states, while the former ones
would qualify more like meson baryon quasibound states or meson baryon
molecules.

   So far, in the light quark section, the dynamically generated baryon
resonances have been found only in the interaction of the octet of stable
baryons with the octet of pseudoscalar mesons in $L=0$, leading to $J^P=1/2^-$
\cite{oller,Oset:2001cn,Jido:2003cb,Garcia-Recio:2003ks} and in the interaction of the
decuplet of baryons with the octet of pseudoscalar mesons in $L=0$, leading to
$J^P=3/2^-$ states \cite{lutz}. 

   The chiral unitary approach is purely a theoretical tool to describe from
scratch the interaction of mesons with baryons. If one uses as input only the
contact Weinberg-Tomozawa interaction between the mesons and baryon as we do,
the interaction is fixed and there is only a free parameter, the cut-off in the
loop, or a subtraction constant in the dispersion relation integral, which is
fitted to a piece of data. However, this cut-off or
subtraction constant should be of natural size \cite{oller}. After that the theory makes predictions for meson
baryon amplitudes and some times a pole appears indicating one has generated a
resonance, which was not explicitly put into the scheme. These are the
dynamically generated resonances. Most of the resonances listed in the PDB can not be
generated in this way indicating they are genuine and not dynamically generated.
Trivial examples of genuine resonances would be the decuplet of baryons to which
the $\Delta(1232)$ belongs. 

With current claims about the $\Theta^+$ pentaquark \cite{Nakano:2003qx} and
the extensive work to try to understand its nature \cite{Hosaka:2003jv,Jaffe:2003sg} (see refs.
\cite{Song:2004bm,Bijker:2004gr} for a list of related references), one is immediately driven to 
test whether such a state could qualify as a dynamically generated resonance from
the $KN$ interaction, but with a  basically repulsive $KN$ interaction from the
dominant Weinberg-Tomozawa Lagrangian this possibility is ruled out.

In view of that, the possibility that it could be a bound state of $K \pi N$ was soon
suggested \cite{Bicudo:2003rw}, but detailed calculations using the same methods 
and interaction that lead to dynamically generated mesons and baryons
indicate that it is difficult to bind that system  with natural size parameters
\cite{Llanes-Estrada:2003us}.  

More recently some new steps have
been done in the chiral symmetry approach introducing the interaction of the
$\Delta$ and the other members of the baryon decuplet with the pion and the
octet partners. In this sense, in \cite{lutz} the
interaction of the decuplet of $3/2^+$ baryons with the octet of pseudoscalar mesons is
shown to
lead to many states that have been associated to experimentally well 
established resonances. Also, in ref. \cite{lutz}
a comment was made that maybe a resonance could be generated with exotic quantum
numbers in the 27 representation of $SU(3)$ . The purpose of the present paper
is to elaborate upon this idea studying the possible existence of a pole in this
exotic channel as well as the uncertainties and stability of the results. 

In the present work we show that the
interaction of the $3/2^+$ baryon decuplet with the $0^-$ meson octet leads
to a state of positive strangeness, with $I=1$ 
and $J^P =3/2^-$, hence, an exotic baryon in the sense that it cannot be
constructed with only three quarks. This would be the first reported case of a dynamically
generated baryon with positive strangeness. However, we study the stability of
the results with reasonable changes of the input parameters and realize that the
results are unstable and the pole disappears within reasonable assumptions. The
situation remains unclear concerning this pole. In view of that we have also
reviewed the situation for the rest of the dynamically generated $3/2^-$
resonances of the model and we find that they are stable and their properties
are quite independent of these changes in the input of the theory. 

\section{Formulation}  
    The lowest order chiral Lagrangian for the interaction of the baryon 
decuplet with the octet of pseudoscalar mesons is given by \cite{Jenkins:1991es}
\be
{\cal L}=-i\bar T^\mu {\cal D}\!\!\!\!/ T_\mu 
\label{lag1} 
\ee
where $T^\mu_{abc}$ is the spin decuplet field and ${\cal D}^{\nu}$ the covariant derivative
given by
\be
{\cal D}^\nu T^\mu_{abc}=\del^\nu T^\mu_{abc}+(\Gm^\nu)^d_aT^\mu_{dbc}
+(\Gm^\nu)^d_bT^\mu_{adc}+(\Gm^\nu)^d_cT^\mu_{abd}
\ee
where $\mu$ is the Lorentz index, $a,b,c$ are the $SU(3)$ indices and $\Gm^\nu$
is the vector current given by
\be
\Gm^\nu=\frac{1}{2}(\xi\del^\nu \xi^\dagger+\xi^\dagger\del^\nu \xi)
\ee
with
\be
\xi^2=U=e^{i\sqrt{2}\Phi/f}
\ee 
where $\Phi$ is the ordinary 3$\times$3 matrix of fields for the pseudoscalar 
mesons \cite{Gasser:1984gg} and $f$ is the pion decay constant, $f=93$ MeV. 
For the $s$-wave interaction some simplifications are possible in 
the algebra of the Rarita-Schwinger fields $T_\mu$
\cite{bookericson}. We write $T_\mu$ as $Tu_\mu$ where $u_\mu$ stands for 
the Rarita-Schwinger spinor which is given by \cite{bookericson,holstein}
\be
u_\mu=\sum_{\lambda, s}{\cal C}(1~\frac{1}{2}~\frac{3}{2}\ ;\ \lambda~s~s_\Delta)\
e_\mu(p,\lambda)\ u(p,s)
\ee
with $e_\mu=(0,\hat e)$ in the particle rest frame, $\hat e$ the spherical
representation of the unit vector $(\lambda=0,\pm 1)$, ${\cal C}$ the Clebsch
Gordan coefficients and $u(p,s)$ the ordinary Dirac spinors
$(s=\pm\frac{1}{2})$. Then eq.~(\ref{lag1}) involves the Dirac matrix elements
\be
\bar u(p\pr,s\pr)\gamma^\nu\ u(p,s)=\delta^{\nu
0}\delta_{ss\pr}+{\cal O}(|\vec p|/M)
\ee
which for the $s$-wave interaction can be very accurately substituted by the
non-relativistic approximation $\delta^{\nu 0}\delta_{ss\pr}$ as done
in~\cite{Oset:1997it} and related works. The remaining combination of the spinors
$u_\mu u^\mu$ involves
\be
\sum_{\lambda\pr, s\pr}\sum_{\lambda, s}
{\cal C}(1~\frac{1}{2}~\frac{3}{2}\ ;\ \lambda\pr~s\pr
~s_\Delta)\ e^*_\mu(p\pr,\lambda\pr) \
{\cal C}(1~\frac{1}{2}~\frac{3}{2}\ ;\ \lambda~s~s_\Delta)\ 
e^\mu(p,\lambda) \ \delta_{ss\pr}=-1+{\cal O}(|\vec p|^2/M^2)~.
\ee 

As one can see, at this point one is already making a nonrelativistic
approximation and consistently with this, the decuplet states will be treated as
ordinary nonrelativistic particles in what follows, concerning the propagators,
etc. However, while making these nonrelativistic assumptions in the Lagrangian
we shall keep the exact relativistic energies in the propagators. This small
inconsistency is assumed in order to find a compromise between simplicity of the
formalism and respecting accurately the thresholds of the reactions and the
exact unitarity.

The interaction Lagrangian for decuplet-meson interaction can then be written in 
terms of the matrix 
\be
(\bar T\cdot T)_{ad}=\sum_{b,c}\bar T^{abc}T_{dbc}
\ee
as
\be
{\cal L}=3iTr\{\bar T\cdot T\,\,\,\Gm^{0T}\}
\label{lag2}
\ee
where $\Gm^{0T}$ stands for the transposed matrix of $\Gm^0$, with  $\Gm^{\nu}$ 
given, up to two meson fields, by
\be
\Gm^\nu=\frac{1}{4f^2}(\Phi\del^\nu\Phi-\del^\nu\Phi\Phi).
\ee 

For the identification of the $SU(3)$ components
of $T$ to the physical states we follow ref. \cite{savage}:

$T^{111}=\Delta^{++}$, $T^{112}=\rth\Delta^{+}$, $T^{122}=\rth\Delta^{0}$,
$T^{222}=\Delta^{-}$, $T^{113}=\rth\Sigma^{*+}$, $T^{123}=\rsix\Sigma^{*0}$,
$T^{223}=\rth\Sigma^{*-}$,  $T^{133}=\rth\Xi^{*0}$,
$T^{233}=\rth\Xi^{*-}$, $T^{333}=\Omega^{-}$.

Hence, for a meson of incoming (outgoing) momenta $k(k\pr)$ we obtain 
 the transition amplitudes, as in \cite{Oset:1997it},
\be
V_{ij}=-\frac{1}{4f^2}C_{ij}(k^0+k^{\pr 0}).
\label{poten}
\ee 

For strangeness $S=1$ and charge $Q=3$ there is only one channel $\Delta^{++}
K^+$ which has $I=2$. For $S=1$ and $Q=2$ there are two channels 
$ \Delta^{++}K^0$ and $\Delta^{+}K^+$ that we call channels 1 and 2, for
which eq. (\ref{lag2}) gives $C_{11}=0$, $C_{12}=C_{21}=-\sqrt{3}$,
$C_{22}=-2$. From these one can extract the transition
amplitudes for the $I=2$ and $I=1$
combinations and we find 
\be
V(S=1,I=2)=\frac{3}{4f^2}(k^0+k^{\pr 0}); ~~~~~V(S=1,I=1)=-\frac{1}{4f^2}
(k^0+k^{\pr 0}).
\label{pot}
\ee
These results indicate that the
interaction in the $I=2$ channel is repulsive while it is attractive in $I=1$.
There is a link to the $SU(3)$ representation since we have the decomposition
\[8\otimes 10=8\oplus 10\oplus 27\oplus 35\]
and the state with $S=1,\ I=1$ belongs to the 27 representation while the $S=1,\
I=2$ belongs to the 35 representation. As noted in \cite{lutz} the
interaction is attractive in the 8, 10 and 27 representations and repulsive in
the 35. Indeed the strength of the interaction in those channels is
proportional to 6, 3, 1 and --3. The attractive potential in the case of $I=1$
and the physical situation are very similar to those of the
$\bar{K}N$ system in $I=0$, where the interaction is also attractive and leads to
the generation of the $\Lambda(1405)$ resonance 
\cite{Kaiser:1995eg,Oset:1997it,oller,Garcia-Recio:2002td}. 
The use of $V$ of  eq. (\ref{poten}) as the kernel of the Bethe Salpeter equation
\cite{Oset:1997it}, or the N/D unitary approach of \cite{oller} both lead to the
scattering amplitude in the coupled channels
\be
t=(1-VG)^{-1}V
\label{LS}
\ee
although in the cases of eq. (\ref{pot}) we have only one
channel for each $I$ state.  In eq. (\ref{LS}), $V$ factorizes on shell
\cite{Oset:1997it,oller} and $G$ stands for the loop function of the meson and baryon
propagators, the expressions for which are given in \cite{Oset:1997it} for a cut off
regularization and in \cite{oller} for dimensional regularization. 

\section{Results and Discussion}

\begin{figure}
\centerline{\includegraphics[width=0.7\textwidth]{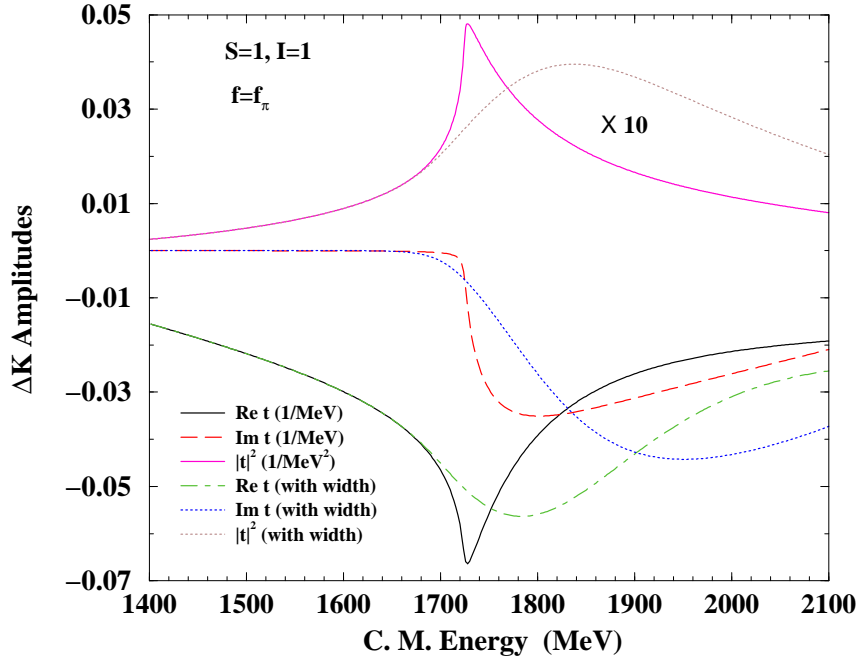}}
\caption{Amplitudes for $\Delta K\rightarrow\Delta K$ for $I=1$ with $f=f_\pi$.}
\label{figI1}
\end{figure}

\begin{figure}
\centerline{\includegraphics[width=0.7\textwidth]{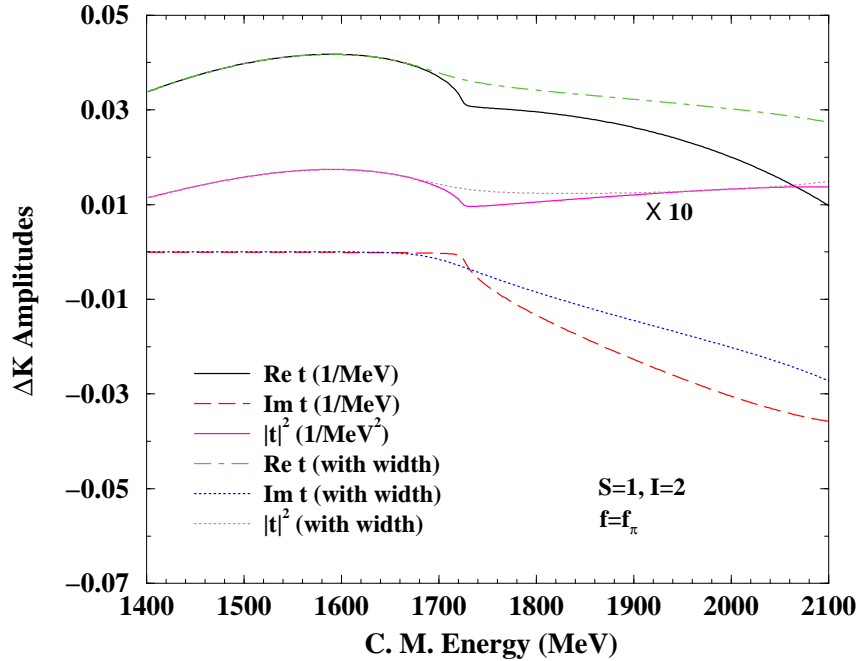}}
\caption{Amplitudes for $\Delta K\rightarrow\Delta K$ for $I=2$ with $f=f_\pi$.}   
\label{figI2}
\end{figure} 

The first thing we have to do is to fix the scale of regularization in the loop
functions $G_{l}$ of eq. (6) of \cite{Oset:2001cn}. The criterion for that is
given in \cite{oller} where dimensional regularization is used and $G_l$ depends
upon a subtraction constant, $a_l$, that should have 'natural size'. Values of 
$a_l$ around $-2$ were found reasonable in \cite{oller} since they are
equivalent to using cut off regularization with $q_{max}$ around 700 MeV
\cite{oller}. In this latter reference, the authors established the
equivalence between the
dimensional regularization and a cut off method in which the
$q^0$ integration in the loops is done analytically and the cut off is put in
the three momentum $\vec q$. Thus, both regularization methods respect the basic
symmetries of the problem. Details on the two methods and  related formulae
used can be seen in ref. \cite{decu_ss} where a general study of the dynamically
generated $3/2^-$ resonances is done.  Here we study in detail the case of
$S=1$,
given the repercussion that such an exotic dynamically generated resonance would
have.

We set up the value of $a_l$ or equivalently $q_{max}$ by fixing the poles for
the resonances which appear more cleanly in \cite{lutz,decu_ss}. They are one resonance in
$(I,S)=(0,-3)$, another one in $(I,S)=(1/2,-2)$ and another one in
$(I,S)=(1,-1)$. The last two appear in \cite{lutz} around 1800 MeV and 1600 MeV and they are
identified with the $\Xi(1820)$ and $\Sigma(1670)$.  We obtain the same results
as in \cite{lutz} using $a_l=-2$ or, equivalently, a cut off $q_{max}=700$ MeV. There are other peaks of the speed plot in 
\cite{lutz}, which we also reproduce,
but they appear just at the threshold of some channels and stick there even when
the cut off is changed. Independently of the meaning of these peaks,
they can not be used to fix the scale of regularization.  
      
  With this subtraction constant we explore the analytical properties of the amplitude for
$S=1$, $I=1$ in the first and second Riemann sheets. Firstly, we see that there is
no pole below threshold in the first Riemann sheet as it would be in the case 
of a bound state. However, if we increase the cut off to 
1.5 GeV (or, equivalently, $a_l=-2.9$ with $\mu=700$ MeV) we find a pole below threshold 
corresponding to a $\Delta K$ bound state. But
this cut off or subtraction constant does not reproduce the position of the resonances discussed above.

  Next we explore the second Riemann sheet. This is done using
dimensional regularization setting the scale
$\mu$ equal to $q_{max}=700$ MeV  and the subtraction constant $a$ to $-2$
and changing
$\bar q_l$ to $-\bar q_l$ in the analytical formula of $G_l$ in 
\cite{Oset:2001cn}. This procedure is equivalent to taking
\be
G^{2nd}=G+2i\,\frac{p_{CM}}{\sqrt{s}}\,\frac{M}{4\pi}
\ee
with the variables on the right hand side of the equation evaluated in the first (physical) 
Riemann sheet. In the above equation $p_{CM}$, $M$ and $\sqrt{s}$ denote the CM 
momentum, the $\Delta$ mass and the CM energy respectively. 
With both methods we find a pole around $\sqrt{s}=1600$ MeV in the second Riemann
sheet. This should have some repercussion on the physical amplitude as we show below. 

The situation in the scattering matrix is revealed in figs. \ref{figI1} and 
\ref{figI2} which show
the real and imaginary parts of the $\Delta K$ 
amplitudes for the case of $I=1$ and $I=2$
respectively. Using the cut off discussed above we can observe  
 the differences between $I=1$ and $I=2$.  For the case of
$I=2$ the imaginary part follows the ordinary behaviour of the opening of a 
threshold,
growing smoothly from threshold. The real part is also smooth, showing
nevertheless the cusp at threshold.
For the case of $I=1$, instead, the strength of the imaginary part is stuck to
threshold as a reminder of the existing pole in the complex plane, growing very
fast with energy close to threshold.  The real part
has also a pronounced cusp at threshold, which is also tied to the same
singularity.  

   We have also done a more realistic calculation taking into account the width
of the $\Delta$ in the intermediate states. For this we use the cut off method
of regularization with $G$ given in \cite{Oset:2001cn} for stable intermediate
particles. The width of the $\Delta$ is taken into account by
adding $-i\Gamma(q^2)/2$ to the $\Delta$ energy, $E_l(\vec q)$, 
in the loop function $G_l$ 
of eq. (6) of ref \cite{Oset:2001cn}
with $\Gamma (q^2)$ given by
\be
\Gamma (q^2)=\Gamma_0\,\frac{q_{CM}^3}{\bar q_{CM}^3}\,\,\Theta(\sqrt{q^2}-M_N-m_\pi)
\ee
where $q_{CM}$ and $\bar q_{CM}$ denote the momentum of the pion (or nucleon)
in the rest frame of the $\Delta$ corresponding to invariant masses
$\sqrt{q^2}$ and $M_\Delta$ respectively. In the above equation
$\Gamma_0$ is taken as 120 MeV.  
The results are also shown in figs. \ref{figI1} and \ref{figI2} and we see that the peaks around
threshold become smoother and some strength is moved to higher energies.  Even
then, the strength of the real and imaginary parts in the $I=1$ are much larger
than for $I=2$. The modulus squared of the amplitudes shows some
peak behavior around 1800 MeV  in the case of $I=1$, while it is small and has no
structure in the case of  $I=2$.

\begin{figure}
\centerline{\includegraphics[width=0.7\textwidth]{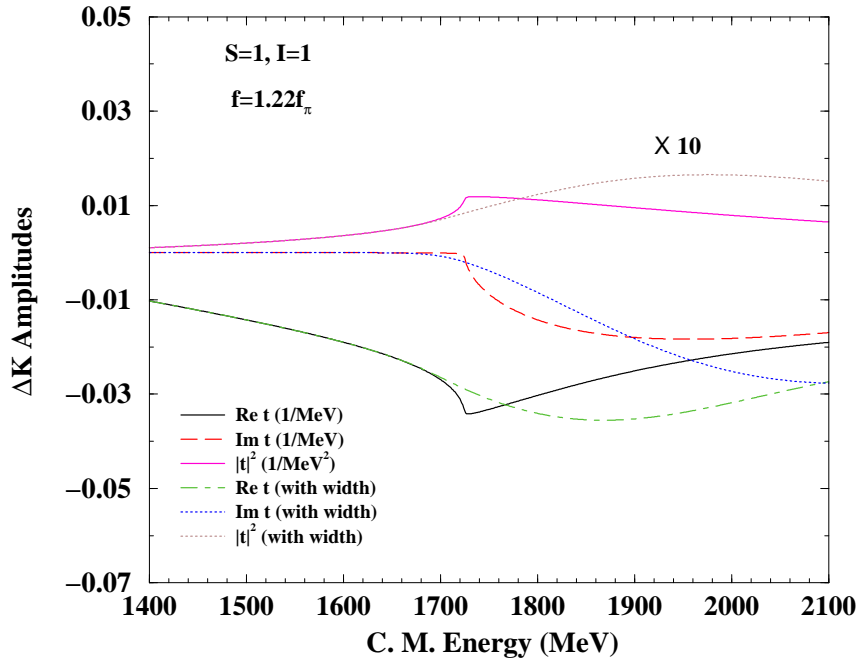}}
\caption{Amplitudes for $\Delta K\rightarrow\Delta K$ for $I=1$ with $f=f_K$.}
\label{newfigI1}
\end{figure}

\begin{figure}
\centerline{\includegraphics[width=0.7\textwidth]{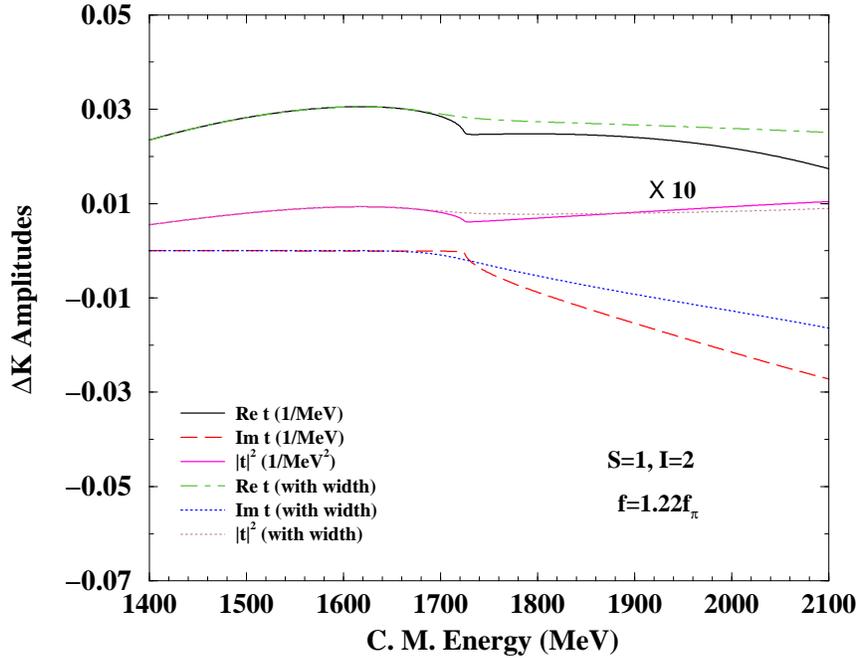}}
\caption{Amplitudes for $\Delta K\rightarrow\Delta K$ for $I=2$ with $f=f_K$.}   
\label{newfigI2}
\end{figure}

The situation in figs. \ref{figI1} and \ref{figI2} is appealing but before
proceeding further we would like to pay some attention to the stability of the
results. So far we have used a unique meson decay constant, the one of the pion,
$f=f_\pi=93$ MeV. One source of $SU(3)$ breaking in the problem comes from the
renormalization of the meson decay constants which leads to different values of
$f$ for the $\pi$, $K$ and the $\eta$ \cite{Gasser:1984gg}. We
thus repeat the calculations using $f_K=1.22f_\pi$. Yet, when doing this, we
would like to change simultaneously the cut off such that we
still obtain the poles for the $\Xi(1820)$ and $\Sigma(1670)$ (which also
involve mainly the $K$ in their coupled channels). We repeat the calculations
with $q_{max}$=800 MeV  and the results are shown in figs. \ref{newfigI1}
and \ref{newfigI2}. We see in fig. \ref{newfigI1} that the cusp effect is very
much diminished with respect to the former set of parameters and the final cross
section is decreased by about a factor of three. Compared to the cross section
for $I=2$, shown in fig. \ref{newfigI2}, the cross section in $I=1$ is still
bigger and grows faster but the effects are certainly less spectacular than
before.

The weak signal in the case of fig. \ref{newfigI1} reflects the fact that that
in this case we do not find a pole in the second Riemann sheet. The interaction,
which is a factor 6 weaker than in the octet case, as we mentioned above, and
barely supported a pole when using $f=f_\pi$, becomes now too weak due to the
$f^{-2}$ behaviour of the kernel (eq. \ref{poten}) and the pole fades away.

In this exploratory investigation we should also mention that if $f$ is
decreased by about \(25\%
\) 
with respect to $f_\pi$ we do find a pole in the first Riemann sheet just below
the $\Delta K$ threshold indicating a bound state, something also mentioned in \cite{lutz}. We have also studied the results making small changes
in the mass of the $K$. Results depend weakly on the $K$ mass but qualitatively
we find that increasing the $K$ mass the interaction becomes stronger (see eq.
\ref{poten}) and it is easier to find the pole, and vice versa.

We are thus in the border line between having and not having a pole, or in
other words, the amplitudes are very sensitive  to changes in the input
parameters. We can not draw strong conclusions in this case since improvements
in the theory could move the balance to one side or the other.

The former comment is in place since a more refined model should also contain
extra channels which have been omitted here. These channels would be states made
of a vector meson and a stable baryon which would also couple in $s$-wave,
the $K^* N$ in the present case. These extra channels are expected to be
relatively unimportant in the case of the other $3/2^-$ dynamically generated
resonances \cite{lutz,decu_ss}, because the Weinberg-Tomozawa interaction is six
times, or three times larger, for the octet or decuplet representations
respectively, than the present one which belongs to the 27 representation.
So, in the present case where we look for the $\Delta K$ pole, the strength
of the $\Delta K$ interaction is rather weak, and the effect of
the other coupled channels and their interaction could alter substantially the
results. We do not have at hand the theoretical tools to study the mixing of 
these
channels and hence it is not possible presently to draw any other conclusions
than the fact that the existence of the $\Delta K$ pole in $I=1$ is rather
uncertain  and a clear answer should wait till better theoretical tools are at
hand or an experiment settles the question.
 
    The next issue concerns the possible experimental reactions that would help
in learning about the $\Delta K$ dynamics. 
The most obvious experiment should be $K^+ p$ scattering which is  
already $I=1$. The state we are generating has spin
and parity $3/2^-$, since the kaon has negative parity and we are working in
$s$-wave in $\Delta K$.  These quantum numbers can only be reached with $L=2$ in 
the $K^+ p$
system.  Thus, the possible resonance should be seen in $K^+ p$ scattering in $d$-waves. 
We estimate that this resonance should have a small effect in $K^+p$ scattering
in $L=2$ based on the experimental fact that the cross section for
$K^+p\rightarrow\Delta K$ is of the order of 1 mb \cite{Giacomelli:1976ig},
while we find here that the $\Delta
K(I=1)$ cross section is of the order of 30--80 mb. The small overlap between 
$K^+p$
 and $\Delta K$ would drastically reduce the effects of the $S=1$, $I=1$ 
 $\Delta K$ state
in $K^+p$ scattering, which could explain in any case why a resonance has never been
claimed in $L=2$ \cite{Hyslop:1992cs}. We have developed a dynamical model for the
$K^+p\rightarrow\Delta K$ overlap and find the conclusions drawn before.
In view of this, we search for other reactions where the existence of the
resonance could eventually be evidenced. Suitable reactions for this would be:
 1) $pp \to \Lambda \Delta^+ K^+$, 2) $pp \to \Sigma^- \Delta^{++}
K^+$, 3) $pp \to \Sigma^0 \Delta^{++}K^0$. In the first case the $\Delta^+ K^+$ state
produced has necessarily $I=1$.  In the second case the $\Delta^{++}K^+$ state has
$I=2$. In the third case the $\Delta^{++}K^0$ state has mostly an $I=1$ component.
A partial wave analysis of these reactions pinning down the $\Delta K$ $s$-wave
contribution would clarify the underlying dynamics of these systems but is
technically involved. Much simpler and still rather valuable would be the information
provided by the invariant mass distribution of 
$\Delta K$, and the comparison of the $I=1$ and
$I=2$ cases. Indeed, the mass
distribution is given by
\begin{equation}
\frac{d \sigma}{dm_{I(\Delta K)}} = C |t_{\Delta K \rightarrow \Delta K}|^2
p_{CM}
\label{sig}
\end{equation}
where $p_{CM}$ is the $K$ momentum in the $\Delta K$ rest 
frame. The mass distribution removing the  $p_{CM}$ factor in eq. (\ref{sig}) 
could eventually
show the broad peak of $|t_{\Delta K \rightarrow \Delta K}|^2$ seen in fig. 1.  Similarly, the ratio of mass
distributions in the cases 3) to 2) or 1) to 2), discussed before, could show 
this behaviour. Similarly, with the help of theoretical calculations of these
reactions at the tree level, the experiment would provide information for the
relative strength of $|t_{\Delta K}|^2$ in $I=1$ compared to $I=2$.
   
    In addition to this test of the mass distribution, one could measure
polarization observables which could indicate the parity or spin of the system
formed, analogously to what is proposed to determine these quantities in
\cite{hosaka,hanhart} or \cite{hyodo} respectively.

On a different note, since we are making a test of stability of the poles we have taken advantage to
see what happens to the poles of the dynamically generated resonances in
\cite{lutz,decu_ss}. We have changed $f$ by $f_K$ in one case or by $1.15f_\pi$
as in \cite{Oset:1997it} to see how much the results change. What we see is that
the poles do not disappear but their positions change. Real parts change by
about 50 MeV on an average, which is well within uncertainties from other
sources and the lack of additional channels \cite{decu_ss}. The widths change in
amounts of the order of 20\(\%
\) except in cases where the shift in mass opens considerably the phase space
available for the decay. However, there is one more significant quantity, the
coupling of the resonance to the different channels, which is calculated from
the residue at the poles. Partial decay widths can be calculated more accurately
using the value of these couplings and the physical mass of the resonances to be
strict with the phase space, as done in \cite{decu_ss}. This exercise served in
\cite{decu_ss} to make a proper identification of the poles found with the
physical resonances. What we observe here is that with the changes in $f$
discussed above, the couplings change by less than 10\(\%
\) on an average and thus the partial decay widths calculated in \cite{decu_ss}
survive the error analysis done here.

This means that the rest of the resonances claimed in \cite{decu_ss} stand on
firm ground and they are quite stable under reasonable changes of the input
parameters. Although fine tuning can be expected from the introduction of extra
channels, the basic features deduced in \cite{decu_ss} should remain unchanged.

\section*{Acknowledgments}
This work is partly supported by DGICYT contract number BFM2003-00856,
and the E.U. EURIDICE network contract no. HPRN-CT-2002-00311.
This research is part of the EU Integrated Infrastructure Initiative
Hadron Physics Project under contract number RII3-CT-2004-506078.

\end{document}